\documentclass[runningheads]{llncs}
\usepackage[T1]{fontenc}
\usepackage{graphicx}
\usepackage{comment}
\usepackage{multirow}
\usepackage{balance}
\usepackage{stfloats}
\usepackage{xcolor}
\usepackage{booktabs}
\usepackage{cite}
\usepackage{amsmath}
\usepackage{comment}

\begin{document}
\title{Enhancing LLM-Based Vulnerability Detection \\via Representation-Aware Logical Ensemble}
\titlerunning{Enhancing LLM-Based Vulnerability Detection via RALE}
%

\author{Francesco Crescenzo Grasso\inst{1,*}\orcidID{0009-0004-1551-1628} \and
Pietro Liguori\inst{1}\orcidID{0000-0001-5579-1696} \and
Roberto Natella\inst{2}\orcidID{0000-0003-1084-4824} \and 
Domenico Cotroneo\inst{3}\orcidID{0000-0001-7103-592X}\\ 
\normalsize *corresponding author}
\authorrunning{F.C. Grasso et al.}
%
\institute{DIETI, University of Naples Federico II, Naples, NA 80125, Italy\\
\email{\{francescocrescenzo.grasso,pietro.liguori\}@unina.it} \and
Computer Science, Gran Sasso Science Institute, L'Aquila, AQ 67100, Italy\\
\email{roberto.natella@gssi.it} \and
Computer Science, University of North Carolina at Charlotte, Charlotte, NC 28223, USA\\
\email{d.cotroneo@charlotte.edu}}

\maketitle              

\begin{abstract}
Automated vulnerability detection in C source code is a long-standing challenge, as traditional static analysis tools often trade precision for scalability and struggle with complex vulnerability patterns. Recently, large language models (LLMs) have shown promising results; yet, most approaches rely on a single code representation, implicitly assuming uniform suitability across vulnerability types.
In this work, we systematically study LLM-based vulnerability detection under multiple code representations, including raw source code, Abstract Syntax Trees (AST), and Control Flow Graphs (CFG). Through extensive empirical analysis, we show that different representations exhibit complementary strengths, with specific views being more effective for particular classes of vulnerabilities. Building on these findings, we propose a Representation-Aware Logical Ensemble (RALE) that integrates LLMs trained on different code representations through explicit logical rules and majority voting. While single-representation models achieve up to 73\% F1-score, our integrated approach reaches 78\% F1-score, with commensurate gains in accuracy, precision, and recall, consistently outperforming individual models. These results highlight the importance of explicitly leveraging complementary code representations when combining LLMs for effective vulnerability detection.

\end{abstract}

\keywords{Software Security \and Vulnerability Detection \and Code Representation Learning \and Ensemble Learning.}

\section{Introduction}
\label{sec:introduction}

The detection of software vulnerabilities represents a key challenge in computer security. Weaknesses such as memory corruption errors and improper resource management continue to be among the most exploited classes of vulnerabilities \cite{mitre2025cwetop25}, motivating the need for automated and reliable detection techniques that can scale to large and complex codebases. Vulnerability detection traditionally relies on static or dynamic analysis. However, static tools tend to produce a high number of false positives \cite{ayewah2007evaluating}, while dynamic tools are limited by execution coverage and
computational costs \cite{dynamic}. Furthermore, both methods do not natively operate at a function-level granularity \cite{survey_static}.

In recent years, data-driven approaches based on deep learning \cite{arewethereyet,zhou2019devign,li2018vuldeepecker,chen2023diversevul,russell} have emerged as a promising alternative to traditional static analysis, aiming to automatically learn vulnerability patterns directly from source code. More recently, advances in large language models (LLMs) have further improved performance in a wide range of code-related tasks, including vulnerability detection, by leveraging large-scale pretraining and contextual representations.

Most of the state-of-the-art approaches process source code primarily as plain text \cite{li2018vuldeepecker,chen2023diversevul,russell}, relying on token-based representations that only implicitly encode program semantics. As a result, crucial structural properties, e.g., syntactic hierarchy and control-flow behavior, are either weakly captured or entirely overlooked. Many studies \cite{arewethereyet,zhou2019devign} attempt to address these limitations by modeling code as graphs, such as Code Property Graphs (CPGs)  or Abstract Syntax Trees (ASTs), explicitly encoding structural and execution-related information. These methods typically focus on a single representation and do not fully exploit the complementary nature of different program views. Consequently, their detection capability remains constrained to the specific aspects emphasized by the chosen representation.

In this work, we explore vulnerability detection through the lens of code representation. Rather than assuming a single optimal view of code, we investigate how different program representations, i.e., raw source code, Abstract Syntax Trees (ASTs), and Control Flow Graphs (CFGs), capture complementary vulnerability-related signals. Using real-world datasets, we systematically train and evaluate LLM-based classifiers on each representation and analyze their behavior at both global and vulnerability-type levels. Building on this analysis, we propose a representation-aware logical ensemble that combines predictions across heterogeneous representations using explicit and interpretable logical rules. The design of the ensemble is guided by security-oriented considerations, favoring decision strategies that mitigate missed vulnerabilities while preserving robustness on non-vulnerable samples.

Our work proposes the following contributions:
\begin{itemize}
    \item We present a comprehensive evaluation of LLM-based vulnerability detection across heterogeneous code representations, showing that raw code, ASTs, and CFGs capture complementary security-relevant information and exhibit distinct specializations across vulnerability types.
    \item We introduce a \textit{Representation-Aware Logical Ensemble} (RALE) that outperforms single-representation models and standard voting schemes, achieving up to 78\% accuracy and F1-score on a real-world dataset.
\end{itemize}

The remainder of the paper is structured as follows. Section \ref{sec:background} elaborates on background concepts related to the target problem. Section \ref{sec:methodology} describes the methodology, including the extraction and generation of representations. Section \ref{sec:experimental_eval} presents the experimental setup and the subsequent evaluation and analysis. Section \ref{sec:lesson} highlights the lessons learned. Section \ref{sec:related} examines related work, while Section \ref{sec:threats} discusses threats to validity. Finally, Section \ref{sec:conclusion} concludes the paper.

\section{Background}
\label{sec:background}

Vulnerability detection aims to identify security flaws in software systems that can be exploited by malicious actors. These weaknesses are commonly categorized according to the Common Weakness Enumeration (CWE) taxonomy \cite{CWE}. In many automated settings, vulnerability detection is formulated as a binary classification task, where the objective is to determine whether a given code fragment is vulnerable or non-vulnerable.

Traditional vulnerability detection relies on static and dynamic analysis. Static tools \cite{Clang:URL, flawfinder} analyze code without execution, often constructing intermediate representations such as ASTs and CFGs; however, they tend to produce a high number of false positives \cite{ayewah2007evaluating}. Dynamic analysis  \cite{lemieux2018fairfuzz,wang2009intscope} observes runtime behavior, offering higher precision in specific scenarios, but it is limited by execution coverage and computational cost \cite{dynamic}. Moreover, most traditional methods do not natively operate at function-level granularity \cite{survey_static}.

Recent research has increasingly explored learning-based approaches to automate vulnerability detection. In particular, large language models (LLMs) have demonstrated strong potential by learning vulnerability patterns directly from large corpora of code. Many existing methods treat source code as plain text \cite{li2018vuldeepecker,chen2023diversevul,russell}, leveraging the ability of neural models to capture statistical regularities in token sequences. While effective, purely token-based representations may overlook structural and semantic properties that are explicitly encoded in program analysis artifacts. To better capture such properties, several works have proposed leveraging graph-based representations of code. Chakraborty \emph{et al.} \cite{arewethereyet} introduced a deep learning approach based on the Code Property Graph (CPG), a unified representation that combines elements of control-flow and data-flow graphs. Similarly, Zhou \emph{et al.} \cite{zhou2019devign} employed CPGs in conjunction with graph neural networks to learn vulnerability patterns. These approaches highlight the importance of structural representations in modeling program behavior.

More generally, different intermediate representations encode distinct aspects of a program. The Abstract Syntax Tree (AST) provides a hierarchical representation of the program’s syntactic structure, explicitly modeling relationships between language constructs, such as expressions and statements. In contrast, the Control Flow Graph (CFG) captures execution semantics by representing possible control transfers between basic blocks of instructions. Each representation emphasizes specific syntactic or semantic properties, suggesting that they may offer complementary perspectives for downstream analysis tasks.

In machine learning, ensemble learning is a well-established paradigm for improving predictive performance by combining multiple models. Ensemble strategies have also been explored in vulnerability detection to improve robustness and predictive stability \cite{Prennsem,ensemble}. The central intuition is that different models may capture different patterns in the data, and their errors are often only partially correlated. By aggregating diverse predictors through strategies such as majority voting, weighted voting, or probability averaging, ensemble methods can reduce variance, mitigate individual biases, and improve robustness and generalization. When models operate on heterogeneous input representations, ensemble techniques can further leverage representational diversity, potentially integrating complementary information captured by distinct views of the same underlying data.

\section{Methodology}
\label{sec:methodology}

Our proposed framework follows a structured process of data generation, representation-specific fine-tuning, and logical ensemble aggregation to maximize vulnerability detection coverage.

\subsection{Data Pipeline}


To support vulnerability detection and further analysis, we adopted \textit{MegaVul}, a large-scale real-world vulnerability dataset for C/C++ introduced by Chao \emph{et al.} \cite{megavul}. It includes 17,380 vulnerable and 322,168 non-vulnerable functions extracted from 992 open-source repositories, covering 169 CWE types. Compared to other datasets \cite{chen2023diversevul,grahn2021analysis}, a key feature is the availability of multiple code representations. In addition to raw source code, it provides node paths for:
\begin{itemize}
    \item \textit{Abstract Syntax Trees} (ASTs), capturing the hierarchical syntactic structure of the function;
    \item \textit{Control Flow Graphs} (CFGs), modeling the execution flow through branches, loops, and decision points.
\end{itemize}
Both representations are generated using the \textit{Joern} \cite{joern_io_Joern_The_Bug_2024} static analysis tool. They enable a systematic comparison of LLM performance across different structural views of the same code.


\vspace{2pt}
\noindent
\textbf{Data Extraction}. Since the graphs are provided in a raw format, an intermediate processing phase was needed to make them suitable for learning models. The extraction process involved parsing the serialized graph outputs to reconstruct the topology of each function. 
For each function, we represent the code as a directed graph $G = (V, E)$, where each node $v \in V$ is characterized by its syntactic label (e.g., $CALL$), operation name (e.g., $<operator>.assignment$), and the corresponding source code snippet. The edge set $E$ encodes structural hierarchy in the AST and execution flow in the CFG. 

All graph representations were converted into adjacency-list format to ensure uniformity across representations and compatibility with LLM inputs.

\begin{table*}[ht]
\centering
\caption{Preprocessing and Abstraction of an AST Node equivalent to a single instruction.}
\label{tab:preprocessing_examples}
\begin{tabular}{@{}lll@{}}
\toprule
\textbf{Feature} & \textbf{Complete Instruction} & \textbf{Structural Instruction} \\ \midrule
\textbf{Raw Node} & \texttt{BinaryOperator 0x14e90c2d8 '<' } & \texttt{BinaryOperator} \\
\textbf{Raw Node (Call)} & \texttt{CallExpr 0x14e925f18 'malloc'} & \texttt{CallExpr} \\
\textbf{Code Snippet} & \texttt{*buffer = malloc(kBuffer...)} & \textit{Removed} \\
\textbf{Metadata} & \texttt{<line:9, col:13>}  & \textit{Removed} \\
\textbf{Addresses} & \texttt{0x14e90bf58} & \textit{Removed} \\ \bottomrule
\end{tabular}
\end{table*}

\vspace{2pt}
\noindent
\textbf{Data Generation}. In addition to the dataset-provided graphs, ASTs were independently generated from the source code using \textit{Clang} \cite{Clang:URL} to ensure consistency, enable alternative representations, and mitigate potential tool-specific biases. While Clang reliably produces high-quality ASTs for function-level snippets, its native CFG extraction assumes access to a complete translation unit with fully resolved includes and macro definitions. Since our experimental setting operates on isolated functions extracted from larger projects,  we rely exclusively on Joern’s CFG extraction pipeline, which is designed to process partial code fragments without requiring full project compilation.


\vspace{2pt}
\noindent
\textbf{Data Preprocessing}. Furthermore, we defined two abstraction levels:
\begin{itemize}
    \item \textit{Complete Representation}: Retains the full semantic context, including operation names and code snippets (e.g., variable assignments and function calls), providing the model with the logic of the implementation.
    \item \textit{Structural Representation}: Strips all identifiers and literals, preserving only the node type and the graph topology to evaluate whether the vulnerability can be detected solely through the "shape" of the code.
\end{itemize}

Table \ref{tab:preprocessing_examples} highlights the abstraction of an AST node, corresponding to a single instruction. In this transformation, the \textit{BinaryOperator} and \textit{CallExpr} types are preserved as the fundamental structural identity, while auxiliary information, such as memory addresses and source code locations, is discarded.

AST and CFG representations were further normalized by removing newline characters and filtering instances outside predefined token-length bounds. 

On the other hand, source code has been lightly preprocessed to remove irrelevant elements related to vulnerability semantics. Specifically, comments (single-line $//$ and multi-line $/* ... */$) and pre-processor directives (e.g., $\#include$, $\#define$) were removed, as they do not affect program behavior. 

\vspace{2pt}
\noindent
\textbf{Data Balancing}. State-of-the-art vulnerability detection often contends with highly imbalanced datasets \cite{chen2023diversevul}, where non-vulnerable samples significantly outnumber security flaws (e.g., a 95\% to 5\% ratio). While this reflects real-world conditions, it poses a significant challenge for supervised learning, making accuracy a misleading metric due to majority-class bias.

To mitigate this discrepancy, we experimented with different dataset splits. As shown in Appendix \ref{app:balancing}, a $50/50$ balanced distribution yielded the most robust results across all metrics. Although this strong undersampling deviates from real-world operational distributions, it prevents models from being overwhelmed by the majority class and isolates the true predictive power of each code representation. Consequently, these findings remain highly valuable, providing clear insights into how representations inherently capture vulnerability patterns when freed from severe distributional bias.

\subsection{Vulnerability Detection}

The following stage of the methodology focuses on model training and evaluation, with the primary goal of assessing the effectiveness of different code representations for vulnerability detection. We consider transformer-based models adapted to a binary classification task, where each C/C++ function is labeled as vulnerable or non-vulnerable.

To systematically evaluate the impact of code representation on model performance, we conducted a cross-comparative analysis involving three distinct architectures and seven functional representations. The representations were categorized into Complete and Structural versions, generated through both Joern and Clang frameworks, and covered three fundamental formats: AST, CFG, and Source Code. This exhaustive setup, except for Clang CFG, yielded a total of 21 independent experiments, ensuring that every model–representation pair was rigorously tested. Such a structured approach allows us to isolate the influence of specific graph topologies and semantic densities, ultimately identifying the most suitable representation for vulnerability detection tasks.

A traditional static analysis tool was included as a baseline to provide a reference point for learned approaches. All experiments were conducted under a unified training and evaluation protocol and assessed using standard classification metrics, including accuracy, precision, recall, and F1-score, ensuring that performance differences are primarily attributable to the chosen representation.
\subsection{Representation-Aware Logical Ensemble (RALE)}

Following the \textit{single-representation} experiments, we conducted a detailed empirical analysis of the results across vulnerability categories. The analysis reveals that different code representations exhibit heterogeneous performance patterns: some representations consistently perform better for a specific class of vulnerabilities while underperforming on others. This variability suggests that no single representation is uniformly optimal and that their predictive behaviors are complementary.

Motivated by this observation, we introduced a Representation-Aware Logical Ensemble strategy, whose pivotal idea is to leverage the complementary strengths of heterogeneous code views. Rather than training models jointly, we aggregated the predictions of independently trained representation-specific detectors through explicit logical rules, preserving modularity and interpretability.



\section{Experimental Evaluation}
\label{sec:experimental_eval}

Experimental evaluation involves comparing fine-tuned models, analyzing results, and presenting the RALE framework.

\subsection{Fine-Tuning Setup}
All experiments were conducted on a workstation equipped with an \textit{Intel 12th Gen Core i7} processor, a \textit{NVIDIA GeForce GTX} 1660 GPU with 6 GB of VRAM, and 32 GB of system RAM. This hardware configuration proved sufficient to support training and inference across all experimental phases, enabling repeated evaluation cycles and consistent comparison of model configurations and preprocessing choices.

Selected models, listed below, are pre-trained on large-scale code corpora and subsequently fine-tuned on AST, CFG, and code vulnerability detection tasks:
\begin{itemize}
\item \textit{CodeT5 220M (Base)} \cite{wang2021codet5} is employed as a reference encoder–decoder model, leveraging span-denoising pre-training and a hybrid architecture that has demonstrated strong performance in both code understanding and vulnerability detection tasks.
\item \textit{CodeBERT 125M (Base)} \cite{feng2020codebert}, an encoder-only model based on RoBERTa, is included to evaluate the effectiveness of rich contextual representations during block-level encoding.
\item \textit{CodeGPT 124M} \cite{radford2019gpt2}, a decoder-only autoregressive model, has been adapted for vulnerability detection via classification heads applied to the final hidden states.
\end{itemize}

To ensure architectural consistency for the classification task, CodeBERT, an encoder-only model, was directly fine-tuned with a classification head. For CodeT5, an encoder–decoder model, we followed the approach established by Wang \emph{et al.} \cite{wang2021codet5}, retaining only the encoder component to extract sequence representations. For CodeGPT, a decoder-only model, based on the GPT architecture, we employed the standard sequence classification variant established by Radford \emph{et al.} \cite{radford2019gpt2}, where prediction is performed using the hidden state of the last non-padding token. 
To prevent overfitting, \textit{early stopping} was applied uniformly across all experiments: training was terminated if the validation loss increased with respect to the previous epoch. As a result, the number of effective training epochs varied slightly across models and representations, while ensuring that each model converged to its best validation performance. Data were split into 88\% for training, 6\% for validation, and 6\% for testing.
\newpage
\subsection{Evaluation of Fine-Tuned LLMs}

We evaluate the effectiveness of different code representations across LLMs using the standard metrics, i.e., accuracy, precision, recall, and F1-score.

\begin{table}[!htbp]
\centering
\small
\setlength{\tabcolsep}{3pt}
\renewcommand{\arraystretch}{0.95}
\caption{Performance of LLMs trained on different code representations}
\label{tab:llm-performance}

\begin{tabular}{lcccccccccccc}
\toprule
& \multicolumn{4}{c}{\textbf{CodeT5}}
& \multicolumn{4}{c}{\textbf{CodeBERT}}
& \multicolumn{4}{c}{\textbf{CodeGPT}} \\
\cmidrule(lr){2-5}
\cmidrule(lr){6-9}
\cmidrule(lr){10-13}
\textbf{Encoding}
& \textbf{Acc} & \textbf{Prec} & \textbf{Rec} & \textbf{F1}
& \textbf{Acc} & \textbf{Prec} & \textbf{Rec} & \textbf{F1}
& \textbf{Acc} & \textbf{Prec} & \textbf{Rec} & \textbf{F1} \\
\midrule

Flawfinder
& -- & -- & -- & --
& -- & -- & -- & --
& 0.53 & 0.56 & 0.53 & 0.46 \\
\midrule

Raw Code
& 0.70 & 0.72 & 0.68 & \textbf{0.70}
& 0.73 & 0.72 & 0.74 & \textbf{0.73}
& 0.71 & 0.73 & 0.69 & \textbf{0.71} \\

CC AST
& 0.62 & 0.61 & 0.63 & 0.62
& 0.66 & 0.68 & 0.64 & 0.66
& 0.65 & 0.64 & 0.66 & 0.65 \\

CS AST
& 0.61 & 0.63 & 0.59 & 0.61
& 0.63 & 0.62 & 0.64 & 0.63
& 0.62 & 0.62 & 0.62 & 0.62 \\

JC AST
& 0.70 & 0.71 & 0.69 & \textbf{0.70}
& 0.71 & 0.71 & 0.71 & \textbf{0.71}
& 0.70 & 0.68 & 0.72 & \textbf{0.70} \\

JS AST
& 0.67 & 0.66 & 0.68 & 0.67
& 0.67 & 0.67 & 0.67 & 0.67
& 0.67 & 0.69 & 0.65 & 0.67 \\

JC CFG
& 0.66 & 0.68 & 0.64 & \textbf{0.66}
& 0.67 & 0.69 & 0.65 & \textbf{0.67}
& 0.68 & 0.70 & 0.66 & \textbf{0.68} \\

JS CFG
& 0.65 & 0.65 & 0.65 & 0.65
& 0.66 & 0.67 & 0.65 & 0.66
& 0.67 & 0.67 & 0.67 & 0.67 \\

\bottomrule
\end{tabular}

\vspace{0.5em}

\parbox{\linewidth}{\scriptsize \textit{Note:} In XY AST or XY CFG, X indicates the frontend (J = Joern, C = Clang), and Y indicates the representation type (S = Structural, C = Complete). Flawfinder is included as a static analysis baseline (independent of LLM encodings).}

\end{table}

As shown in Table \ref{tab:llm-performance}, across all models, raw source code consistently achieves the highest overall performance, confirming that lexical and semantic information is highly informative for vulnerability detection. Among structured representations, Joern Complete AST and Joern Complete CFG provide the strongest results, closely matching raw code and outperforming other structural or Clang-based variants. In contrast, representations that remove semantic or execution-related details show noticeable performance drops.

Our empirical observations suggest that performance trends are primarily representation-driven rather than strictly model-specific, highlighting the intrinsic discriminative power of raw code and complete AST/CFG structures. Nevertheless, it's worth noting that while no significant performance bias is observed between the different architectures, CodeBERT \cite{feng2020codebert} mostly achieves the best overall results. This confirms that an encoder-only architecture remains the most effective choice for discriminative tasks such as vulnerability classification. Overall, raw code, Joern Complete AST, and Joern Complete CFG emerge as the best-performing representations for their respective categories, motivating their use in the subsequent representation-aware ensemble.

Importantly, these results raise the question of whether the high-performing representations identify the same vulnerable functions or complementary subsets. Investigating this overlap is crucial to inform the design of the subsequent representation-aware ensemble, where combining complementary predictions can further improve detection performance.

\subsection{Performance Across Vulnerability Macro-Categories}

\begin{figure*}[htbp]
\centering
\centerline{\includegraphics[width=0.9\textwidth]{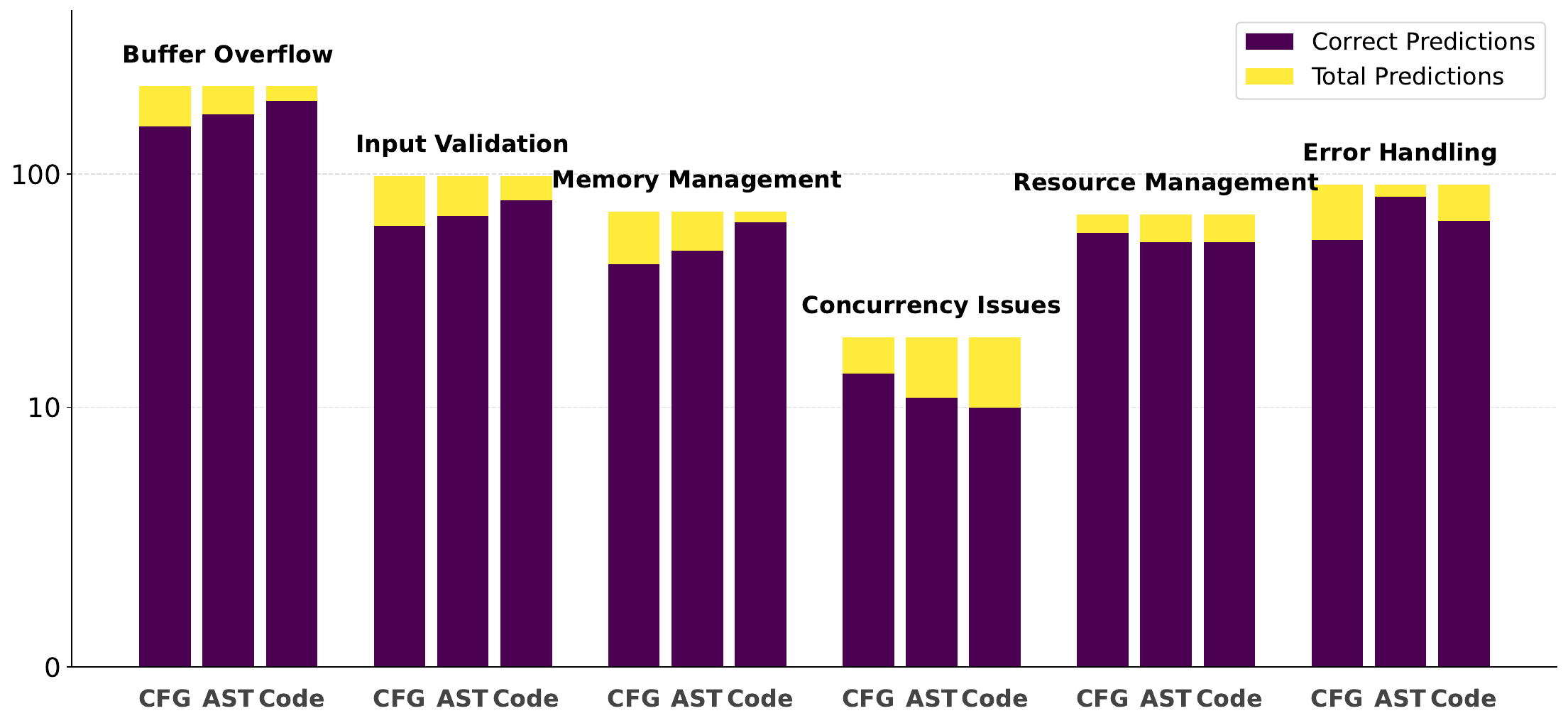}}
\caption{Log-scaled distribution of correct over total predictions of the test set across different representations and selected macro vulnerability categories reported in Appendix \ref{app:map}}
\label{fig:analysis}
\end{figure*}

In this section, we analyze how models trained on different code representations perform across types of vulnerable functions. Figure \ref{fig:analysis} shows a log-scale histogram of correct predictions over the total for each representation and macro-category. To avoid an excessive number of categories, CWEs were grouped into macro-categories, e.g., \textit{CWE-125: Out-of-bounds Read} \cite{cwe125} falls under \textit{Buffer Overflow}; \textit{CWE-20: Improper Input Validation} falls under \textit{Input Validation and Injection}. Full details are provided in Appendix \ref{app:map}.  

We observe that:
\begin{itemize}
    \item \textit{Raw Source Code} performs best on \textit{Buffer Overflow}, \textit{Input Validation and Injection}, and \textit{Memory Management}, as these categories rely heavily on semantic details such as variable usage, pointer operations, and data manipulation patterns, which are most explicitly preserved in raw code.
    \item \textit{CFGs} excel at detecting \textit{Resource Management} and \textit{Concurrency Issues} due to their explicit modeling of execution paths, branching, and loops, which are critical for reasoning about resource lifetimes and race conditions.
    \item \textit{ASTs} show improved performance on \textit{Error Handling and Logic Issues} since their hierarchical structure emphasizes syntactic constructs and control relationships, aiding the detection of incorrect branching, missing checks, or improper exception handling.
\end{itemize}

In summary, raw code, ASTs, and CFGs capture complementary aspects of program behavior: raw code provides rich semantic and syntactic context, ASTs emphasize structural and logical organization, and CFGs highlight execution dynamics. This complementarity explains the observed variation in CWE-level performance and motivates a representation-aware ensemble, which leverages the strengths of each representation to achieve more robust and comprehensive vulnerability detection.

\begin{figure*}[htbp]
\centering
\centerline{\includegraphics[width=0.9\textwidth]{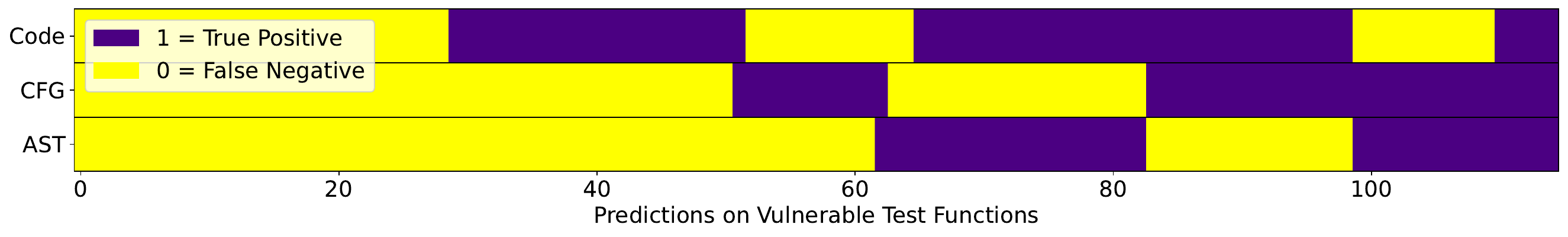}}
\centerline{\includegraphics[width=0.9\textwidth]{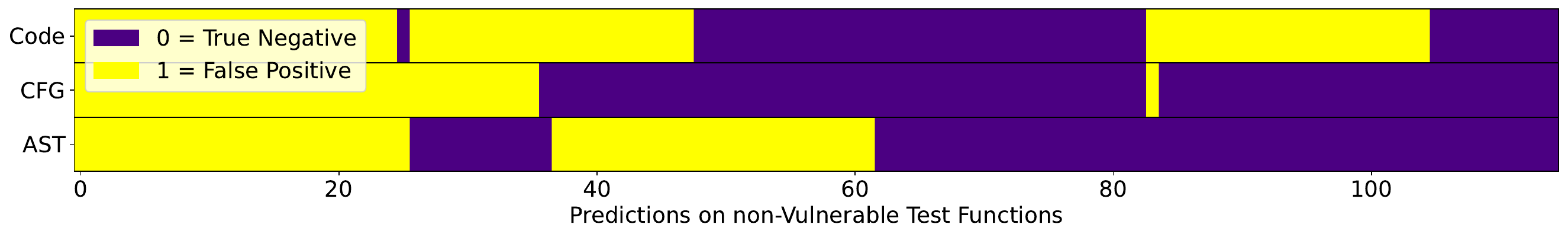}}
\caption{Overlap and complementarity of prediction errors across different representations on a half of the test set where discrepancies between predictions are most pronounced}
\label{fig:complementary}
\end{figure*}

\subsection{File-Level Complementarity Analysis}

A deeper analysis of model predictions at the file level reveals substantial disagreement among models, indicating that each captures only partially overlapping vulnerability signals. We perform a \textit{complementary analysis}, examining whether errors made by one representation are corrected by others. In this context, complementary means that models do not fail on the same examples. For instance, if raw code, AST, and CFG consistently make identical errors, an ensemble would offer no benefit. The key question is whether vulnerabilities missed by one model are correctly identified by another.

Figure \ref{fig:complementary} emphasizes this behavior through a binary heatmap of vulnerable (top) and non-vulnerable (bottom) functions. For clarity, we focus on half of the test set where discrepancies between predictions are most pronounced.

The analysis shows that the raw source code model exhibits a higher false positive rate compared to AST- and CFG-based models, while AST achieves the lowest false positive rate overall. However, this apparent disadvantage is offset by the raw code model’s superior performance on vulnerable samples, frequently detecting cases missed by other representations. Such complementary behavior indicates that simple model selection is suboptimal and motivates combining predictions through logical gates to improve robustness.

\subsection{Logical Ensemble Evaluation}

\begin{figure*}
\centering
\centerline{\includegraphics[width=0.9\textwidth]{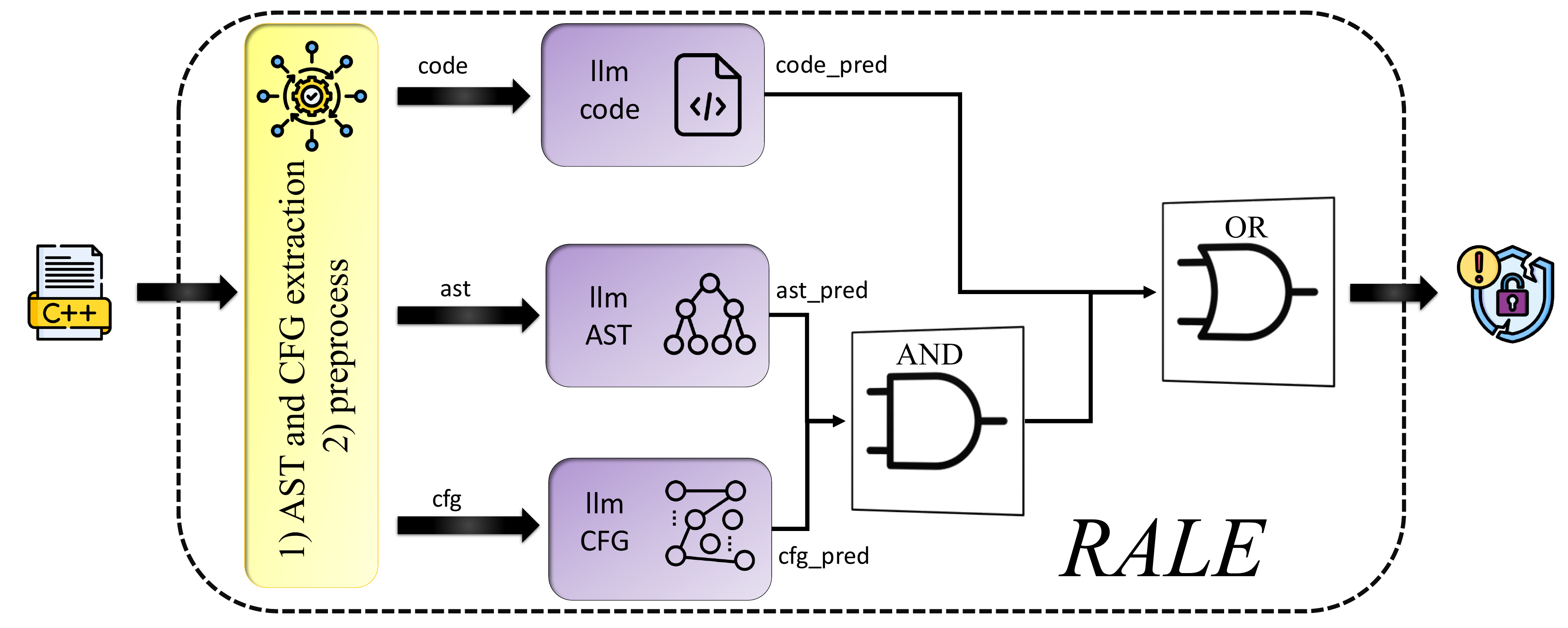}}
\caption{RALE architecture: integration of LLMs trained on Raw Code, AST, and CFG through a representation-aware logical ensemble governed by the boolean expression $(\text{AST} \land \text{CFG}) \lor \text{Code}$}
\label{fig:RALE}
\end{figure*}



As a first aggregation strategy, we adopt a \textit{majority voting} mechanism, i.e., 2-out-of-3, where a function is classified as vulnerable if at least two of the three models (raw code, AST, CFG) predict a vulnerability. This approach reduces the impact of individual model errors, particularly false positives on non-vulnerable samples. Conceptually, this mechanism introduces democratic decision-making, where no single representation dominates.

However, a closer inspection reveals that many false negatives of the raw code model are corrected by AST or CFG. In security-critical contexts, false negatives are more costly than false positives, as undetected vulnerabilities pose a higher risk than over-flagged safe functions. This observation motivates the introduction of a bias in favor of the raw code predictions. 

    To preserve the raw code model’s strong recall, we implement a \textit{hybrid decision rule}. Specifically, if the raw code model predicts a function as vulnerable, this prediction is taken as final. In all other cases, the standard 2-out-of-3 majority voting is applied, which is equivalent to a logical OR between the raw code model and the majority voting system. This hybrid approach combines the high recall of the raw code with the improved false-positive control of the AST and CFG.

\begin{table}[htbp]
\centering
\caption{Performance of ensemble strategies on unseen code}
\label{tab:ensemble-performance}
\begin{tabular}{c c c c c}  
\toprule
\textbf{Representation / Ensemble} & \textbf{Acc.} & \textbf{Prec.} & \textbf{Rec.} & \textbf{F1} \\
\midrule
CodeGPT on JC CFG  & 0.68 & 0.70 & 0.66 & 0.68 \\
CodeBERT on JC AST  & 0.71 & 0.71 & 0.71 & 0.71 \\
CodeBERT on Code & 0.73 & 0.72 & 0.74 & 0.73 \\ 
Majority Voting & 0.75 & 0.75 & 0.75 & 0.75 \\
\textbf{RALE} & 0.78 & 0.77 & 0.79 & \textbf{0.78} \\
\bottomrule
\end{tabular}
\end{table}

Table \ref{tab:ensemble-performance} reports performance using the Ensemble Mechanisms. Majority voting yields a consistent improvement of approximately 2\% across evaluation metrics compared to the raw code baseline, while RALE provides the largest performance gains, with overall improvements of approximately 5\% in accuracy, precision, recall, and F1-score relative to the raw code baseline.

Figure~\ref{fig:RALE} illustrates the RALE architecture. For clarity, the figure presents a simplified version of the ensemble logic using boolean algebra. 
\begin{gather}
    \label{eq:rale}
    (\text{2-out-of-3}) \lor \text{Code} \\ 
    ((\text{AST} \land \text{CFG}) \lor (\text{AST} \land \text{Code}) \lor  \\
    \lor (\text{Code} \land \text{CFG})) \lor \text{Code} \nonumber \\
    (\text{AST} \land \text{CFG}) \lor \text{Code}
\end{gather}
The original weighted ensemble logic (1) can be expressed as (2). By first applying the associativity of $\lor$ and then the absorption law $X \lor (X \land Y) = X$, the expression (2) simplifies to (3). This compact form clarifies the decision rule: a vulnerability is flagged if the Code model detects it, or if both the AST and the CFG models agree. Moreover, it reduces logical complexity and computational overhead compared to heavier ensemble architectures.
\section{Lesson Learned}
\label{sec:lesson}
Our systematic study highlights three fundamental lessons for effective vulnerability detection:

\vspace{2pt}
\noindent
(i) \textit{Complementarity of Code Views}: Since different code representations (e.g., AST, CFG, raw source) highlight distinct structural and semantic properties, models trained on these diverse views capture fundamentally different vulnerability patterns. This suggests that a "one-size-fits-all" representation limits the detection capability of LLMs.

\vspace{2pt}
\noindent
(ii) \textit{Simplicity vs. Effectiveness}: Leveraging multiple representations through a logical ensemble (RALE) proves to be as simple as it is effective. This approach offers a powerful trade-off, achieving high detection performance (up to 78\% F1-score), without the prohibitive computational complexity often associated with more opaque or deeply nested hybrid architectures.

\vspace{2pt}
\noindent
(iii) \textit{Practical Deployment}: In real-world settings, we position RALE as a semi-automated human assistant, given the inevitable trade-offs in precision and recall. Rather than completely replacing manual review, the framework should be leveraged as an early-warning tool that provides initial suggestions on where vulnerabilities might reside.

\section{Related Work}
\label{sec:related}

Traditional approaches rely on static and dynamic analysis. Dynamic analysis \cite{lemieux2018fairfuzz,wang2009intscope} observes program execution to detect vulnerabilities, offering high precision in specific scenarios; however, it is constrained by limited execution coverage and high computational costs. Static analysis \cite{Clang:URL, flawfinder} examines code without execution, often constructing intermediate representations such as ASTs and CFGs. Despite their effectiveness, both approaches have limitations: dynamic analysis requires extensive manual configuration and scales poorly to large codebases, whereas static analyzers tend to produce high numbers of false positives \cite{ayewah2007evaluating}. Moreover, most traditional methods do not natively operate at function-level granularity \cite{survey_static}.

Recent studies have explored data-driven approaches to automate vulnerability identification. Large Language Models (LLMs) have shown strong potential by learning vulnerability patterns directly from large code corpora. Existing methods typically treat code as plain text \cite{li2018vuldeepecker,chen2023diversevul,russell}, leveraging neural networks’ ability to capture statistical regularities in token sequences. In addition, some approaches rely on synthetic code \cite{russell} as training data; however, prior work \cite{arewethereyet} has shown that synthetic samples often fail to reflect real-world vulnerability patterns, limiting their effectiveness in practical settings. Among them, Li \emph{et al.} \cite{li2018vuldeepecker} use \textit{code gadgets} for deep-learning-based vulnerability detection. Chen \emph{et al.} \cite{chen2023diversevul} train token-level models achieving high accuracy but limited F1-score, while Russell \emph{et al.} \cite{russell} learn deep feature representations to identify vulnerabilities.
Several works emphasize the benefits of structural code representations. Chakraborty \emph{et al.} \cite{arewethereyet} employ Code Property Graphs (CPGs), combining control-flow and data-flow information for deep-learning models. Similarly, Zhou \emph{et al.} \cite{zhou2019devign} use CPGs with graph neural networks to capture vulnerability patterns. These studies highlight that different structural representations can capture complementary aspects of program behavior.

In vulnerability detection, ensembles have been applied to enhance robustness and stability \cite{Prennsem,ensemble}. Wang \emph{et al.} \cite{Prennsem} propose PreNNsem, a heterogeneous ensemble operating on sample vectors. Cho do Xuan \emph{et al.} \cite{ensemble} combine a Code Understanding (CU) module with a Vulnerability Encoder (VE) based on embeddings.
While prior work has explored either single-representation LLMs or ensemble strategies, most approaches implicitly assume uniform suitability across vulnerability types. In contrast, our work systematically investigates LLMs over multiple code representations (Raw Code, AST, CFG) and explicitly integrates them via a representation-aware logical ensemble (RALE). This allows us to exploit complementary strengths of each representation, prioritize high-precision components (e.g., Code), and achieve improved detection performance, particularly reducing false negatives.

\section{Threats To Validity}
\label{sec:threats}


We identify several limitations of our study:

\vspace{2pt}
\noindent
(i) \textit{Empirical Scope}: Our findings are grounded in the specific datasets and experimental settings adopted. While these provide a diverse set of examples, results—including the relative performance of code representations—may vary when applied to different programming languages, proprietary codebases, or distinct vulnerability types.

\vspace{2pt}
\noindent
(ii) \textit{Model Selection}: Although we utilize state-of-the-art architectures like CodeBERT \cite{chen2023diversevul}, the choice of backbone affects validity. While our results confirm the superiority of encoder-only architectures for this classification task, using larger generative models or different hyperparameter optimizations might lead to different performance baselines.

\vspace{2pt}
\noindent
(iii) \textit{Test Set Distribution}: During the validation phase, we systematically experimented with different undersampling ratios to determine the optimal training balance. However, our final evaluation was conducted on balanced test sets. While this standard practice ensures a controlled environment for comparing code representations, it does not fully reflect real-world vulnerability detection scenarios, where vulnerable code is naturally highly imbalanced, and the actual false positive rate could be higher.
\section{Conclusion}
\label{sec:conclusion}


Our work highlights the value of representation-aware model integration, showing that carefully designed logical ensembles can significantly improve the effectiveness and reliability of automated vulnerability detection. By leveraging the complementary strengths of source code, AST, and CFG representations, the RALE framework achieves a 5\% improvement in F1-score over single-representation models. Furthermore, a key finding of our study is the efficacy of the hybrid decision rule: by prioritizing the raw code model’s recall while utilizing a 2-out-of-3 majority voting for consensus, we successfully minimize critical false negatives without the typical explosion of false positives. 
\section*{Acknowledgments}
\label{sec:Ack}
This work was supported by the European Union – NextGenerationEU under the PNRR (Piano Nazionale di Ripresa e Resilienza), Mission 4, Component 2, Investment 'I.3.4 Borse PNRR transizioni digitali e ambientali', through Ministerial Decree no. 118 (02-03-2023), and the MUR Department of Excellence 2023 - 2027 grant at GSSI (Gran Sasso Science Institute). The authors would like to express their gratitude to the former student, Luigi Scotto Rosato. 


\appendix
\section{Balancing Test}
\label{app:balancing}

In Table \ref{tab:balancing_impact}, preliminary tests using a \textit{Random Forest classifier} trained on Megavul \cite{megavul} demonstrate that a $50/50$ distribution yields the most robust results across all performance metrics. This balancing strategy ensures that the model assigns equal importance to both classes, preventing the optimization process from being dominated by the majority "non-vulnerable" class, thus enhancing \textit{sensitivity}, i.e., the true positive rate, towards critical security flaws.

\begin{table}[ht]
\centering
\small
\setlength{\tabcolsep}{5pt}
\renewcommand{\arraystretch}{1.1}

\caption{Impact of Dataset Distribution on Model Performance}
\label{tab:balancing_impact}

\begin{tabular}{lcccc}
\toprule
\textbf{Distribution (NV/V)} & \textbf{Acc} & \textbf{Prec} & \textbf{Rec} & \textbf{F1} \\
\midrule

95\% / 5\% (Imbalanced) & 0.95 & 0.00 & 0.00 & 0.00 \\
70\% / 30\%             & 0.78 & 0.52 & 0.45 & 0.48 \\
\textbf{50\% / 50\% (Balanced)} & \textbf{0.68} & \textbf{0.65} & \textbf{0.71} & \textbf{0.68} \\

\bottomrule
\end{tabular}
\end{table}

\section{CWE to Macro-categories Mapping}
\label{app:map}

We grouped individual Common Weakness Enumerations (CWEs) into eight distinct macro-categories, as shown in Table~\ref{tab:cwe-macro}, to evaluate model performance across functionally similar vulnerability types.

\begin{table}[ht]
\centering
\small
\setlength{\tabcolsep}{5pt}
\renewcommand{\arraystretch}{1.15}

\caption{Mapping of CWE Categories to Macro-Categories}
\label{tab:cwe-macro}

\begin{tabular}{p{4.6cm}p{7cm}}
\toprule
\textbf{Macro-Category} & \textbf{Included CWEs} \\
\midrule

Buffer Overflow & CWE-119, CWE-125, CWE-787, CWE-120, CWE-122 \\
Memory Management & CWE-415, CWE-416, CWE-401 \\
Resource Management & CWE-400, CWE-399, CWE-835, CWE-770, CWE-772 \\
Input Validation and Injection & CWE-20, CWE-22, CWE-189, CWE-190 \\
Information Exposure & CWE-200, CWE-254 \\
Concurrency Issues & CWE-362 \\
Security Features & CWE-264, CWE-287 \\
Error Handling and Logic Issues & CWE-617, CWE-369, CWE-476, CWE-674 \\
\bottomrule
\end{tabular}
\end{table}

\bibliographystyle{splncs04}
\bibliography{biblio}

\begin{thebibliography}{8}
\bibitem{ref_article1}
Author, F.: Article title. Journal \textbf{2}(5), 99--110 (2016)

\bibitem{ref_lncs1}
Author, F., Author, S.: Title of a proceedings paper. In: Editor,
F., Editor, S. (eds.) CONFERENCE 2016, LNCS, vol. 9999, pp. 1--13.
Springer, Heidelberg (2016). \doi{10.10007/1234567890}

\bibitem{ref_book1}
Author, F., Author, S., Author, T.: Book title. 2nd edn. Publisher,
Location (1999)

\bibitem{ref_proc1}
Author, A.-B.: Contribution title. In: 9th International Proceedings
on Proceedings, pp. 1--2. Publisher, Location (2010)

\bibitem{ref_url1}
LNCS Homepage, \url{http://www.springer.com/lncs}, last accessed 2023/10/25
\end{thebibliography}

\end{document}